\documentclass[a4paper,10pt]{article}
\usepackage[utf8]{inputenc}
\usepackage{booktabs}
\usepackage{graphicx}
\usepackage{subfigure}
\usepackage{rotating}
\usepackage{placeins}
\usepackage{pdfpages}

\title{Modules for Automated Validation and Comparison of Models of Neurophysiological and Neurocognitive Biomarkers of Psychiatric Disorders:
ASSRUnit - A Case Study} 
\author{Christoph Metzner, Tuomo M\"aki-Marttunen, \\ Bartosz Zurowski and Volker Steuber}

\begin{document}

\maketitle

\begin{abstract}
The characterisation of biomarkers and endophenotypic measures has been a central goal of research in psychiatry over the last years. While most of this research has focused on the identification 
of biomarkers and endophenotypes, using various experimental approaches, it has been recognised that their instantiations, through computational models, have a great potential to help us understand
and interpret these experimental results. However, the enormous increase in available neurophysiological and neurocognitive as well as computational data also poses new challenges. 
How can a researcher stay on top of the experimental literature? How can computational modelling data be efficiently compared to experimental data? How can computational modelling most 
effectively inform experimentalists? Recently, a general scientific framework for the generation of executable tests that automatically compare model results to experimental observations, 
SciUnit, has been proposed. Here we exploit this framework for research in psychiatry to address the challenges mentioned above. We extend the SciUnit framework by adding an experimental
database, which contains a comprehensive collection of relevant experimental observations, and a prediction database, which contains a collection of predictions generated by computational models. 
Together with appropriately designed SciUnit tests and methods to mine and visualise the databases, model data and test results, this extended framework has the potential to greatly facilitate 
the use of computational models in psychiatry. As an initial example we present ASSRUnit, a module for auditory steady-state response deficits in psychiatric disorders.   
\end{abstract}

\section{Introduction}
Psychiatric nosology, for centuries widely untouched by findings from (clinical) neuroscience is at the beginning of a transformation process \cite{Friston2017} 
towards an interactive evolution of diagnostic and biological categories. This change of focus stems from the hope
that biomarkers and endophenotypic measures show a better correspondence with genetic 
alterations identified by large genome-wide association studies \cite{Meyer2006} and promises to more readily shed light on the mechanisms underlying
these disorders and to facilitate the discovery of novel therapeutic interventions \cite{Siekmeier2015}.
Naturally, a lot of effort has been put into the translation of these measures into practice using human studies \cite{Perlis2011}
as well as animal models \cite{Markou2009}.

Computational approaches also have gained significantly more attention over the last years and this has led to the emergence of 
'Computational Psychiatry'
as a novel multidisciplinary and integrative discipline (see for example \cite{Montague2012,Wang2014,Friston2014,Corlett2014,Stephan2014,Adams2016}).
This emergence can be attributed to three main factors: First, the above mentioned increase in experimental studies has provided 
a wealth of neuroscientific (including neurochemical, molecular, anatomic, and neurophysiological) data which are essential to build  computational models. Second, methodological and infrastructural advances,
such as the  various atlases, databases and online tools from the Allen Brain Institute (\textit{http://brain-map.org/}) or the BRAIN initiative 
(\textit{https://www.braininitiative.nih.gov/}),
have made it possible to analyze and process this enormous amount of data. Third, the increase in computing power of high performance
computers as well as standard personal computers has made it possible (and affordable) to build and use models of increasingly high computational 
complexity. Therefore, the rapid growth of the field of computational psychiatry comes as no surprise.
However, in order to fully exploit the potential that computational modeling offers, we have to identify systemic weaknesses
in current approaches and take a look at other disciplines that use computational models (and have used them for much longer than psychiatry)
and even look at disciplines, like software development, which face similar challenges.

At the core of computational modeling lies the concept of validation, i.e. the rigorous comparison of model predictions 
against experimental findings. Furthermore, for a model to be useful and provide a true contribution to knowledge, the validation has to 
use sound criteria and the experimental observations need to sufficiently characterize the phenomenon the model tries to reproduce.
Hence, in order to develop a computational model scientists need to have an in-depth understanding of the current, relevant experimental data, 
the current state of computational modeling in the given area and the state-of-the-art of statistical testing, to choose the appropriate criteria
with which the model predictions and experimental observations will be compared \cite{Gerkin2013,Sarma2016}. In a field where both the number of experimental 
and computational
studies grows rapidly, as is the case for psychiatry, this becomes more and more impracticable.
Furthermore, the increase in modeling and experimental studies has made it harder for reviewers not only to judge whether a new model adequately 
replicates the full range of experimental observations but also how it compares to competing models. Again, also reviewers need an 
in-depth knowledge of the modeling
and experimental literature as well as profound statistical knowledge.
Finally, since computational modeling tries to generate predictions which can be experimentally tested, experimental neuroscientists must be able 
to extract and assess predictions from a rapidly growing body of computational models, a task which is also becoming more and more impracticable.

The problems described above are not unique to the field of computational psychiatry but occur in all scientific areas that use computational models.
Furthermore, building a computational model is in the end a software development project of sort. Omar et al. \cite{Omar2014}
have therefore
proposed a framework for automated validation of scientific models, SciUnit, which is based on unit testing, a technique commonly used in software 
development.
SciUnit addresses the problems mentioned above by making the scope (i.e. the set of observable
quantities that it can generate predictions about) of the model explicit and by allowing its validity (i.e. the extent to which its predictions
agree with available experimental observations of those quantities) to be automatically tested \cite{Omar2014}.

In this paper, we  propose to adopt this framework for the computational psychiatry community and to collaboratively build
common repositories of computational models, tests, test suites and tools.
As a case in point, we have implemented a Python module (\textit{ASSRUnit}) for auditory steady-state response (ASSR) deficits in schizophrenic patients, which are based on 
observations from
several experimental studies (\cite{Krishnan2009,Vierling2008,Kwon1999})and we demonstrate
how existing computational models (\cite{Metzner2016,Beeman2013,Vierling2008,Metzner2017}) can be validated against these observations and compared with each 
other.


\section{The SciUnit Framework}
\begin{figure}
\includegraphics[width=\textwidth]{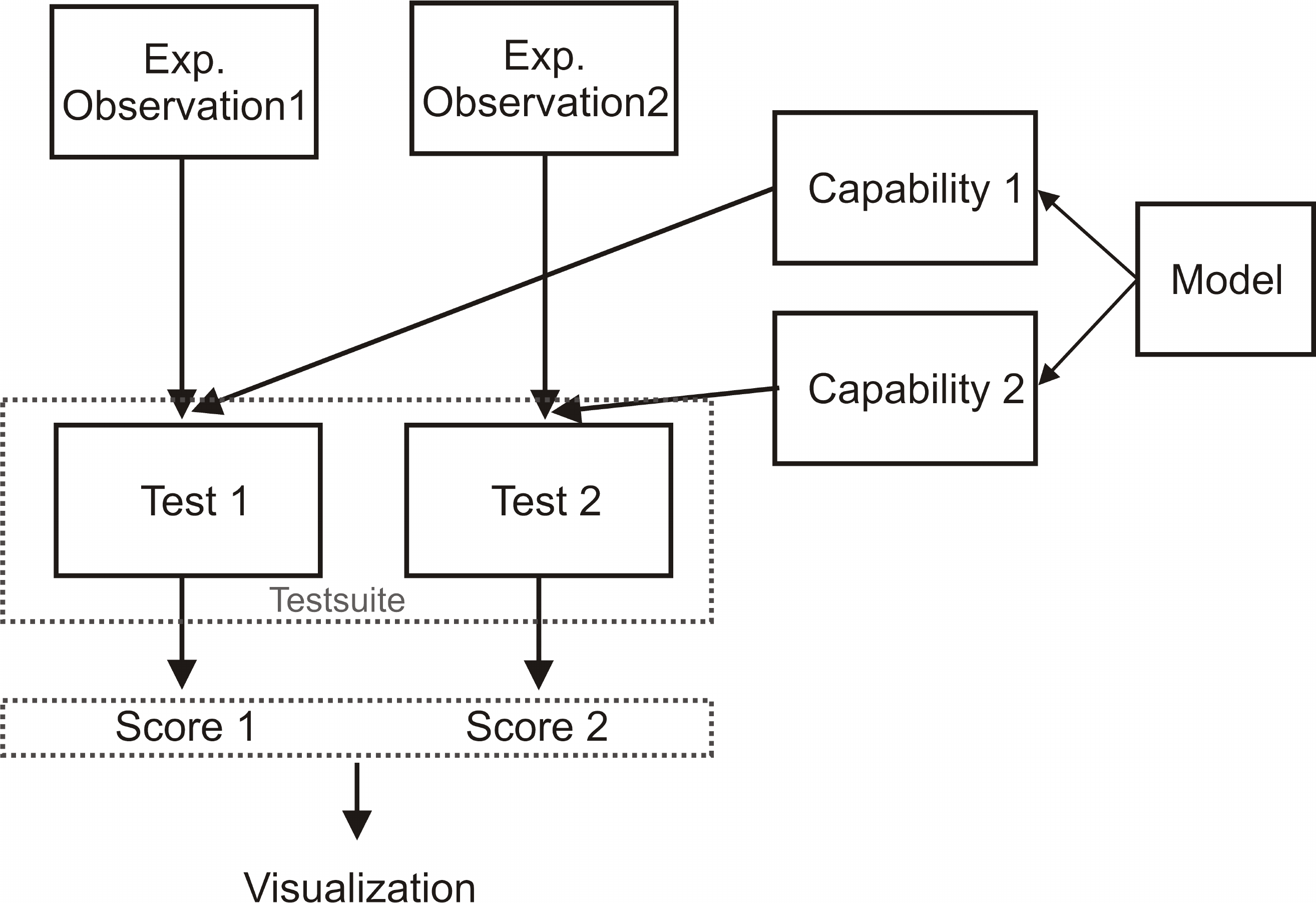}
\caption{Schematic representation of the SciUnit framework. Models can be tested against experimental observations using specific tests. 
These tests incorporate an experimental observation and interface with the model through capabilities. Tests can be grouped into so-called test suites. The execution of a test 
produces a score, which describes how well the model captures the experimental observations. SciUnit also provides methods to visualize the resulting score(s), for example in a table.}
\label{Fig:SciUnit-Scheme}
\end{figure}

The module we present here is based on the general SciUnit framework for the validation of scientific models against 
experimental observations \cite{Omar2014} (see Figure \ref{Fig:SciUnit-Scheme}).

In SciUnit models declare and implement so-called capabilities, which the validation tests then use to 
interact with those models. By a capability of the model, we mean the ability of the model to describe certain biological phenomena that are possible to assess using physical quantities.
Furthermore, the declaration
and implementation of capabilities are separated, which allows to test two different models that share the 
\textit{same capabilities} on the \textit{same experimental observations} using
the \textit{same test}. Tests then take the model, use its capabilities to generate data and compare these data to the 
experimental observations which are linked to the test
and create a score. This score, which can simply be a Boolean (pass/fail) or another more complex score type, 
describes if and to which extent the model data and the experimental observation(s) match.

Before we describe the actual implementations of capabilities, models, tests and scores in our framework for 
ASSRs in schizophrenia,
we first start with a summary of the experimental observations we included in the database and 
then we describe the computational models which were realized.

\section{The \textit{ASSRUnit} Module}
\begin{figure}
\includegraphics[width=\textwidth]{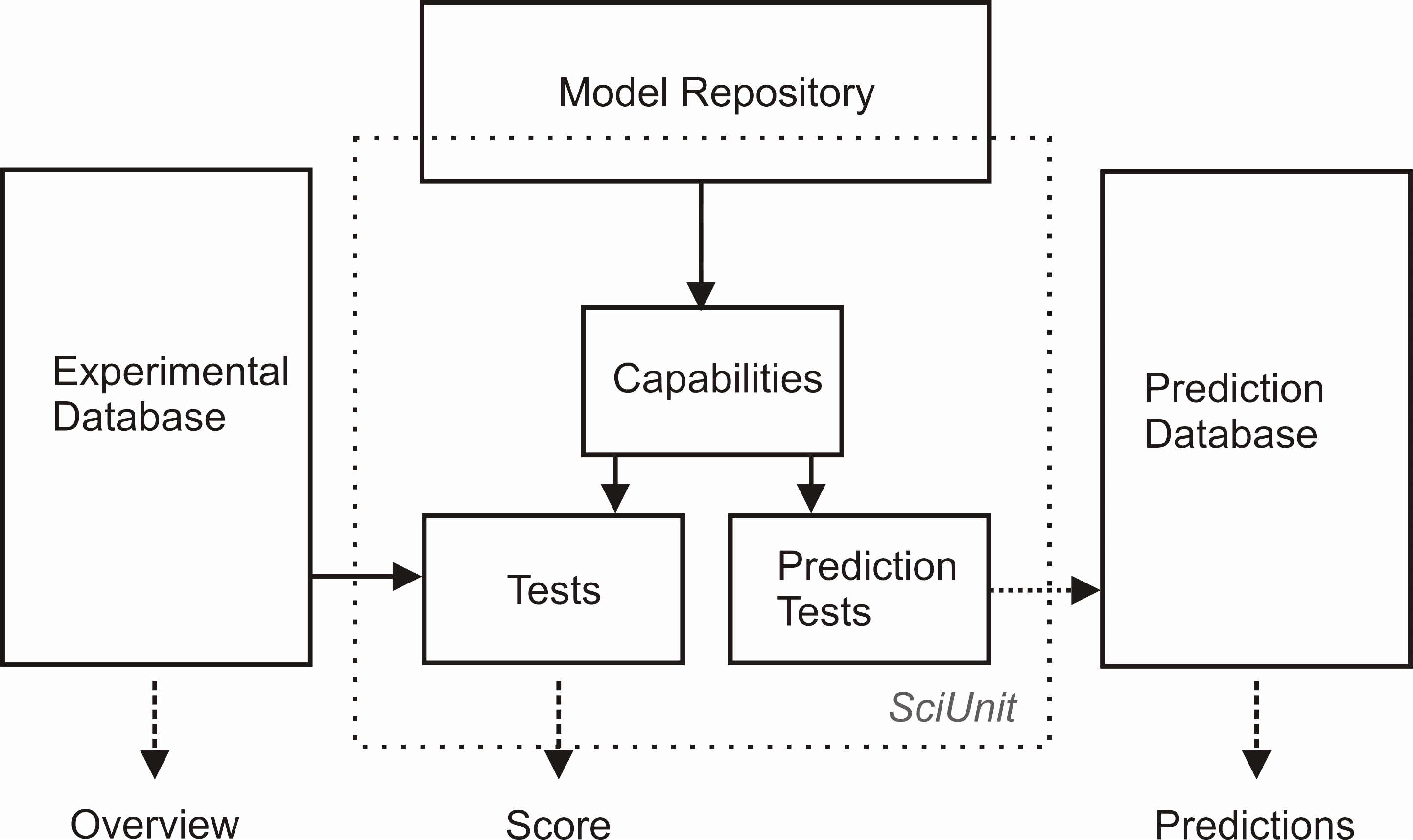}
\caption{Schematic representation of the proposed framework highlighting the three main functions: 1) Overview of experimental observations. 2) Validation of computational models. 3) Creation of a predictions database. At its core
lies the SciUnit module, which provides the infrastructure for the automated validation of the computational models. In particular, through a set of suitable tests, the computational models can be compared against experimental observations
queried from the experimental database. Another set of tests, the so-called prediction tests, are then employed to extract predictions from the computational models, thus populating the predictions database. }
\label{Fig:Scheme}
\end{figure}

The structure of the ASSRUnit module proposed here is shown schematically in Figure \ref{Fig:Scheme}. As outlined earlier, there are three main functionalities the proposed module aims to provide: 1) To provide a simple way of getting an overview
of the experimental literature, 2) To provide an easy and flexible way to automatically test computational models against experimental observations, 3) To provide an automated way of generating predictions from computational
models. Functionality 1 is fully covered by the experimental database and its methods to query the database and visualize the results. Functionality 2 is provided by linking both the experimental database as well as the 
computational models to the SciUnit tests that cover the relevant experimental obervations. The only action required from the user is, if the computational model has not yet been included into the model repository of the module, 
to provide an interfacing Python class for the model which implements all the required capabilities. Note that the model itself does not have to be written in Python, it only has to be executable from shell. Once the model is
included, the SciUnit framework allows for automated testing and the visualization methods provided in the proposed module allow for a comprehensive and clear presentation of the results. Functionality 3 can be achieved by
a set of SciUnit tests and capabilites that, instead of covering experimental observations, cover experiments that have not yet been performed. By running the computational models with these tests, 
the module can be used to generate new predictions from the models, which can then be used to populate a prediction database similar to the experimental database. The module is available on GitHub: 
\textit{https://github.com/ChristophMetzner/ASSRUnit}

\subsection{Experimental Observations Database}

In patients suffering from schizophrenia oscillatory deficits in general and ASSR deficits in particular have been extensively studied using electroencephalography (EEG) and 
magnetoencephalography (MEG) 
(e.g.  \cite{Kwon1999,Vierling2008,Krishnan2009,
Light2006,Zhang2016,Hamm2015,Brenner2003,Spencer2009b,Spencer2008,Spencer2012,OConnell2015, Mulert2011}). Here, we focus on three of these studies looking at
entrainment deficits in the gamma and beta range. Kwon et al. \cite{Kwon1999} used a click train paradigm to study ASSRs at $20$, $30$, and $40$\,Hz in schizophrenic
patients using EEG and found a prominent reduction of power at the driving frequency for $40$\,Hz drive, an increase in power at the driving frequency during $20$\,Hz drive and 
no changes for $30$\,Hz drive. Furthermore, they found small changes of power at certain harmonic/subharmonic frequencies, namely, an increase of power at $20$\,Hz for $40$\,Hz
drive and a decrease of power at $40$\,Hz for $20$\,Hz drive. Vierling-Claassen et al. \cite{Vierling2008} reproduced these findings using the same paradigm with MEG. 
Krishnan et al. \cite{Krishnan2009} used a slightly different paradigm, which employed amplitude-modulated tones instead of click trains, and tested a wide range of driving 
frequencies from $5$ to $50$\,Hz. They found reduction of power at the driving frequency in the gamma range (i.e. at $40$, $45$ and $50$\,Hz) and no changes at other frequencies.
Furthermore, they did not find any changes of power at harmonic or subharmonic frequencies. 

The experimental database is realized as a nested Python dictionary, with an entry for each study included.
Each study entry consists of two entries, which describe the study observations, one in a quantitative way and the other in a qualitative way.
We have included the qualitative description because often either computational models do not allow for a strict quantitative comparison with 
experimental data or publications of experimental studies do not provide enough detail on the results, and in these cases, only a qualitative comparison is possible. 

\begin{table}[h!]
\centering
\label{Tab:Experiments} 
\caption{Summary of ASSR deficits in schizophrenic patients in the three studies considered here ($\downarrow$: sign. lower in patients, $\uparrow$: sign. higher in patients, $-$: no sign. difference between controls and patients). 
Since Kwon et al. \cite{Kwon1999} and Vierling-Claassen et al. \cite{Vierling2008}
produced the same results, they are combined here. The tests included in the \textit{ASSRUnit} module are based on this table. Note that Krishnan et al. \cite{Krishnan2009} tested more driving frequencies than the ones shown in the table. 
The table only shows measures that are common to all three studies.}
\begin{tabular}{lccccccc}\toprule
& \multicolumn{3}{c}{Fundamental}& \phantom{a} & Harmonic & \phantom{a} & Subharmonic \\
\cmidrule{2-4} \cmidrule{6-6} \cmidrule{8-8}
Drive& $40$\,Hz  & $30$\,Hz  & $20$\,Hz&  & $20$\,Hz & &$40$\,Hz \\
\midrule 
Kwon/Vierling & $\downarrow$& -& $\uparrow$& &$\downarrow$ & & $\uparrow$\\ \addlinespace

Krishnan & $\downarrow$& -& -& &- & &- \\
\bottomrule
\end{tabular}
\end{table}

Together with the database, \textit{ASSRUnit} provides basic methods to query and visualize the content of the database. These methods include commands to retrieve all studies or observations in the database and a method
to display an overview of the results for the whole database or for certain studies or observations. Finally, the meta-data associated with each study (for example, the number of participants, the modality, the patient group, etc.)
can also be retrieved and displayed.

\subsection{Prediction Database}
The prediction database is also implemented as a nested Python dictionary. Similar to the experimental observation database, methods that retrieve and visualize the content of the database are included in \textit{ASSRUnit}.

\subsection{Models, Capabilities, Tests and more}

\paragraph{Models}
In order to demonstrate the flexibility of the proposed framework, we included three different neural models of ASSR deficits.

The first model is based on a biophysically detailed model of primary auditory cortex by Beeman \cite{Beeman2013}. It has recently been used to study 
ASSR deficits by our group \cite{Metzner2016}. The model was implemented using the neural simulator GENESIS \cite{Bower1992,Bower1998}. Not only is this model
a good example of a biophysically detailed model of ASSR deficits, its inclusion also demonstrates how models that are not written in Python can be used.

The second model is a reimplementation of the model of Beeman in NeuroML2, a simulator-independent markup language to describe neural network models developed by the
NeuroML project \cite{Cannon2014}, which is featured in the open source brain model database \cite{Gleeson2012}. We included this model to demonstrate the ability of the proposed framework
to incorporate state-of-the-art tools and databases for the design, implementation and simulation of network models.

The last model we included is the simple model presented by Vierling-Claassen et al. \cite{Vierling2008}. The model is a simple network of two populations of theta neurons. 
We reimplemented the model in Python (for more details on the model and the replication see \cite{Metzner2017}).
The model
was included first of all to demonstrate that the framework is not limited to biophysically detailed models but can also be used with simpler, more abstract models.
Additionally, the inclusion of the model demonstrates the simplest way of including a model, implementing the model in Python. This might not be the most common scenario, but 
since it is the simplest, we included it here.

We do not discuss the models in more detail here, since they have been described elsewhere \cite{Beeman2013,Metzner2016,Vierling2008,Metzner2017}. Furthermore, our focus lies on the framework with which to use, validate and compare
models not on the models themselves.

The three models mentioned above are included into the SciUnit framework by wrapper classes that implement the necessary capabilities and make the models available to the tests.
One important thing to note here is that, since we are dealing with models of neurofunctional deficits found in individuals with a particular disorder, a 'model' as used in the module always means two configurations of a computational 
model, one representing the control configuration and one the disorder configuration. Therefore, all wrapper classes take two sets of parameters as an argument describing the necessary parameters
for the two configurations, respectively.
In addition to the standard model classes, we also implemented a second version of the model classes, which simulates a certain number $n$ of simulations, instead of a single one, 
where each simulation differs in background noise. This allows for assessing the robustness of the results.

\paragraph{Capabilities}
Table \ref{Tab:Experiments} summarizes the experimental observations included in the module at this stage. All observations are similar in nature: the power value of the EEG/MEG 
at a certain frequency in response to auditory entrainment at a certain frequency. Therefore, the only capability necessary for a model to produce output that can be compared to these observations
is a method that produces the power at a certain frequency X of a simulated EEG/MEG signal in response to drive at a frequency Y. This capability, \textit{ProduceXY}, is included in ASSRUnit and all models
must implement it.

\paragraph{Tests and Scores}
The five tests we implemented, examine the five observations summarized in Table \ref{Tab:Experiments} individually. Furthermore, we implemented one 
prediction test, which tests 10\,Hz power at 10\,Hz drive.
For the sake of simplicity, the test scores implemented so far are simple Boolean scores, indicating whether a model output fails or passes a test, that is whether the difference 
between model output for the control and the 'schizophrenia-like' network
matches the experimental observation. In case of the model classes implementing sets of outputs, simply the mean difference is compared to the experimental observations.
For the prediction test we have chosen a RatioScore instead of a Boolean, which
returns the ratio of the power for the 'schizophrenia-like'  configuration and the power for the control configuration.

\paragraph{Visualization, Statistics, Additional Data}

In addition to the main features of the SciUnit framework for the analysis and comparison of the models, we use the fact that SciUnit allows to pass additional data, beyond the test scores, to provide
a class that offers tools for the visualization of the results. This class includes functions to display the test results in a table, plot the results from a set of model outputs as 
a box plot, and perform and visualize a student's t-test of the differences between control and 'schizophrenia-like' networks.

Next, we describe three different use cases, which show how the proposed module can be used for different purposes by experimentalists, modelers and reviewers.

\subsection{Use Case I: Overview of the Experimental Literature}
\begin{figure}
\includegraphics[width=\textwidth]{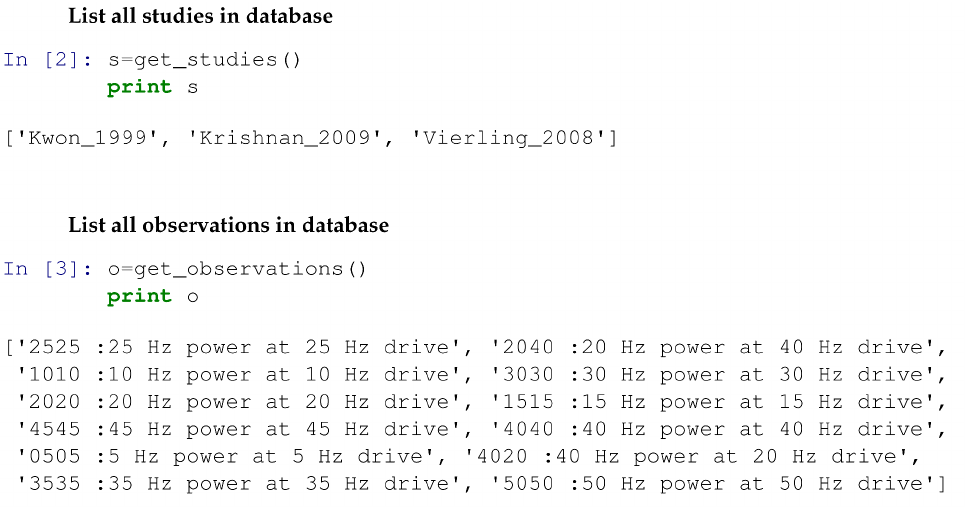}
\caption{Display all studies  and all observations included in the database}
\label{Fig:ListStudies}
\end{figure}

The first use case demonstrates how the experimental database can be used to get a comprehensive overview of the current experimental literature related to a neurophysiological or neurocognitive biomarker, in our case
ASSR deficits in patients suffering from schizophrenia. Figure \ref{Fig:ListStudies} shows that with two
simple commands one can retrieve the names of all studies and all observations present in the database. These names will have to be used for all further queries of the database.

\begin{sidewaysfigure}[ht]
\includegraphics[width=\textwidth]{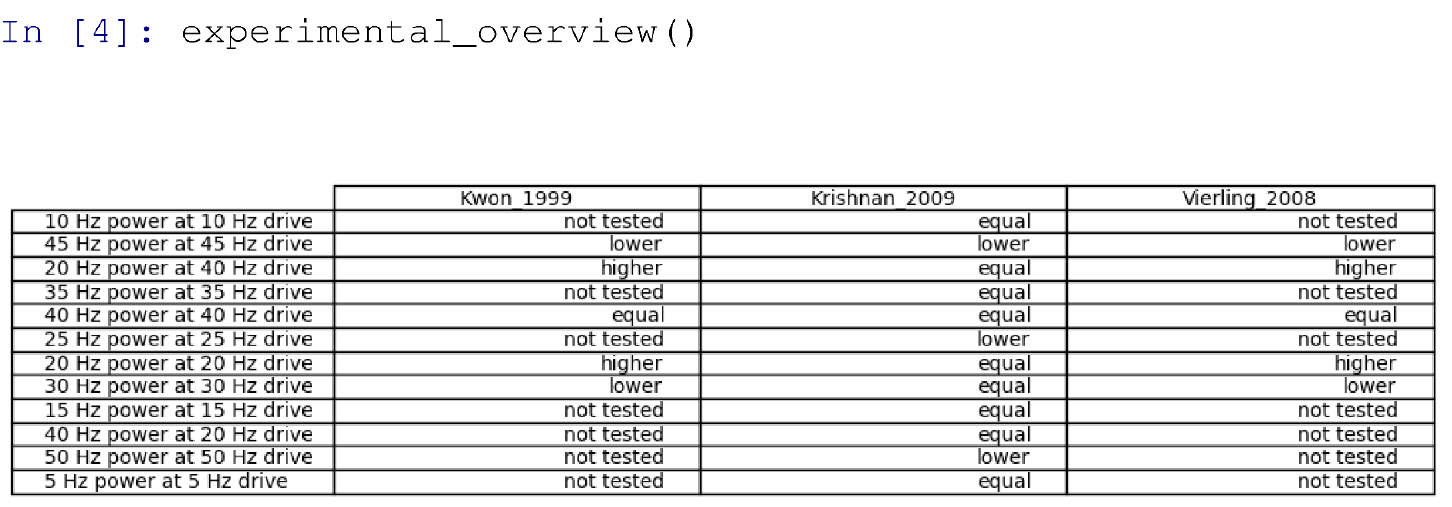}
\caption{Overview of the observations in the experimental literature.  The command \textit{experimental\_overview} prints a table summarizing the results for all studies and all observations in the database. 
 Note that by default the qualitative study results are presented. This can be changed to the quantitative results
setting the parameter \textit{entrytype} to \textit{Full}.}
\label{Fig:ExpOverview0}
\end{sidewaysfigure}

\begin{figure}
\includegraphics[width=\textwidth]{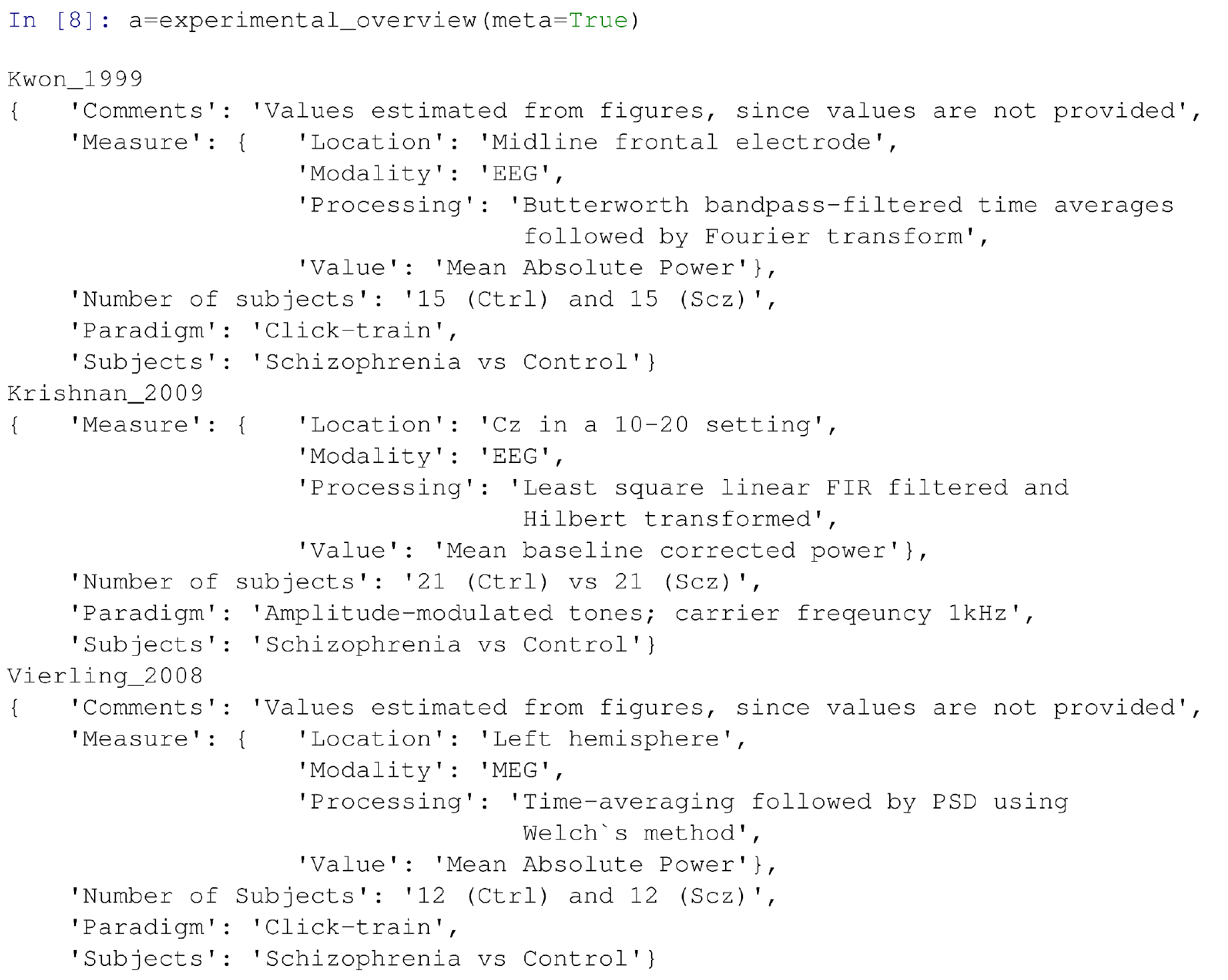}
\caption{ By setting the \textit{meta} flag to \textit{True}, additional information on the studies are displayed.}
\label{Fig:ExpOverview1}
\end{figure}

\begin{figure}
\includegraphics[width=\textwidth]{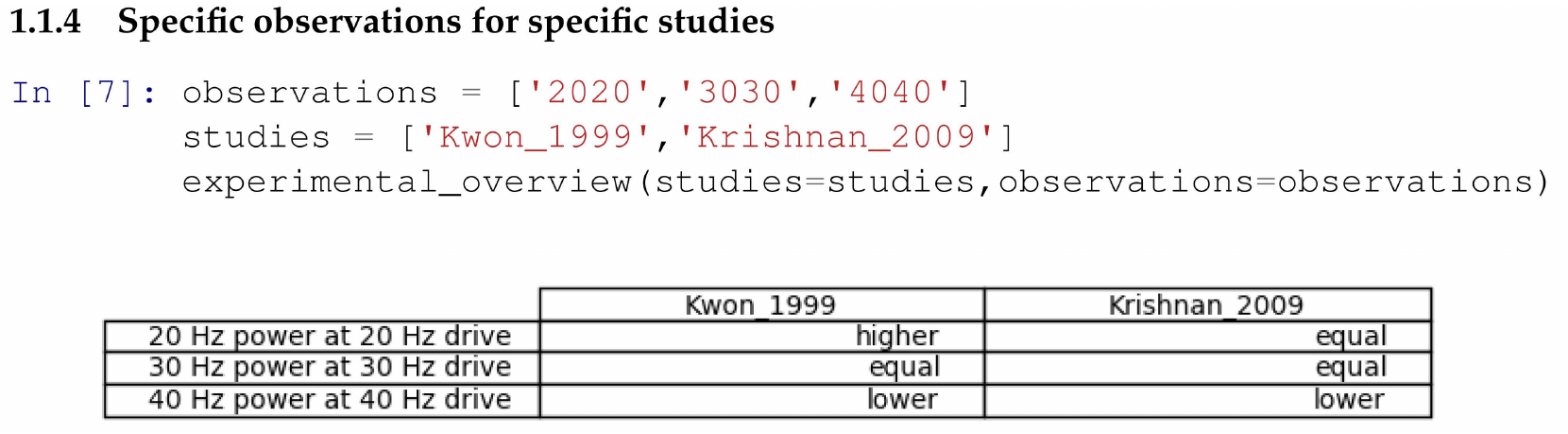}
\caption{The \textit{experimental\_overview} command allows for querying for specific studies and observations using the names retrieved with the \textit{get\_studies} and \textit{get\_observations} commands.}
\label{Fig:ExpOverview2}
\end{figure}

Figure \ref{Fig:ExpOverview1} a) then shows how to get a complete overview of all observations of all studies in the database. As we can see in Figures \ref{Fig:ExpOverview0} \ref{Fig:ExpOverview1} b), simply adding the parameter \textit{meta=true},
to the command, will additionally output the meta-data associated with each study. This contains information on the subjects, modality etc.
The overview command presents the data in a simple table and can be used to see which studies provided 
which observation and what the results were. However, as we can already see for our small demonstration database containing only three studies, this is likely to become big and therefore hard to fully grasp.
By explicitly stating the studies and/or the observations one is interested in, one can reduce the complexity of the table and get a clear and simple overview, as depicted in Figure \ref{Fig:ExpOverview2}. Note that in the
examples, we have only used the qualitative description of the observations, the same functionality also applies to the quantitative descriptions. The functionality described here, along with
more examples, can be explored in an accompanying Python notebook (\textit{Example\_Experimental\_Database.ipynb} in \textit{https://github.com/ChristophMetzner/ASSRUnit/Code/}).

This simple querying functionality allows the user to get a quick, clean and comprehensive overview of the experimental literature, to identify observations that are supported by many studies (see in our case the reduction of
gamma power for stimulation at gamma frequency) but also to detect controversial findings. Furthermore, the display of the associated meta-data allows to check for example whether identified common observations extend over
different modalities and post-processing techniques, and also whether controversial findings might be explained by differences in the experimental setup or other related aspects. In the future, it will also be possible 
to look at more than one database and compare the same observations across different patient groups to highlight commonalities and differences between disorders.

\subsection{Use Case II: Model Comparisons}

While our first use case only exploited the experimental database, we now show the additional benefits of joining experimental and modeling data. 

\begin{figure}
\subfigure[Create model instances]
{\includegraphics[width=0.9\textwidth]{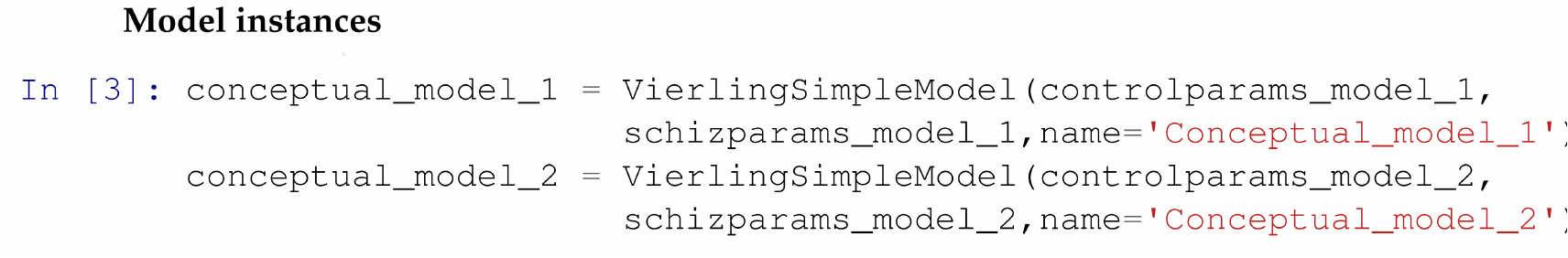}}
\subfigure[Create tests]
{\includegraphics[width=0.9\textwidth]{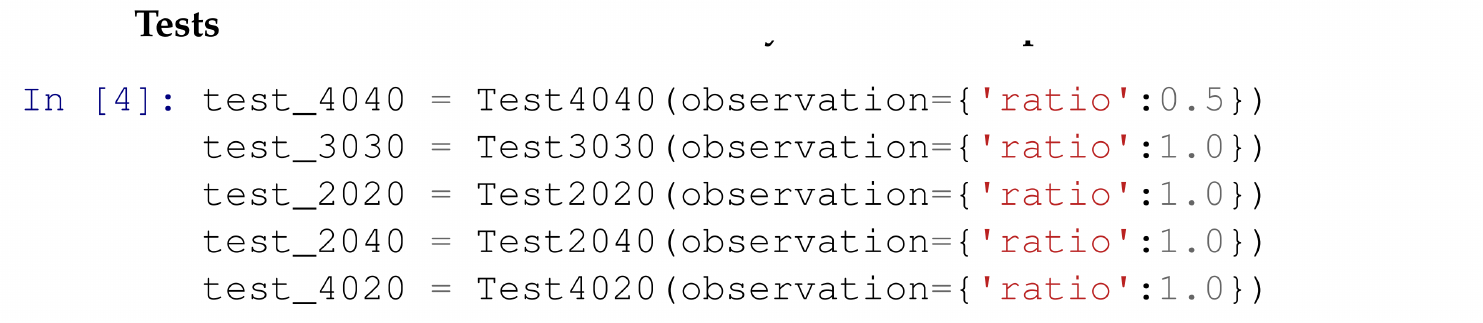}}
\subfigure[Create a testsuite and run models against it]
{\includegraphics[width=0.9\textwidth]{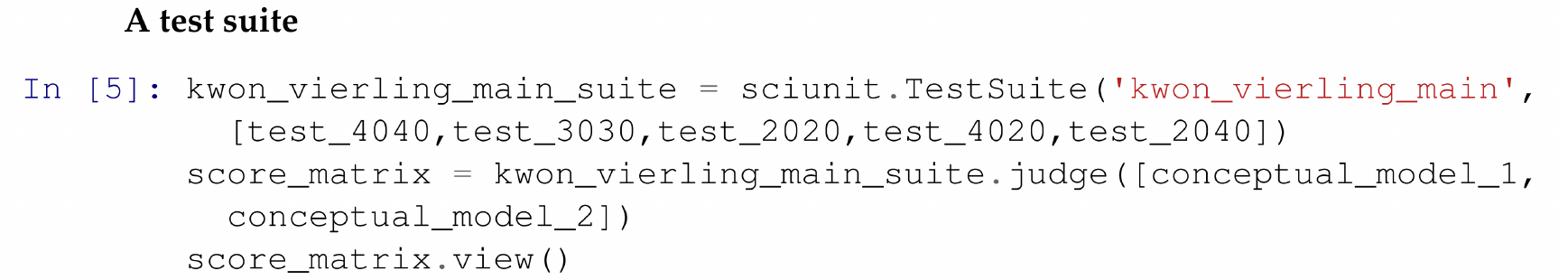}}
\subfigure[Display comparison]
{\includegraphics[width=0.9\textwidth]{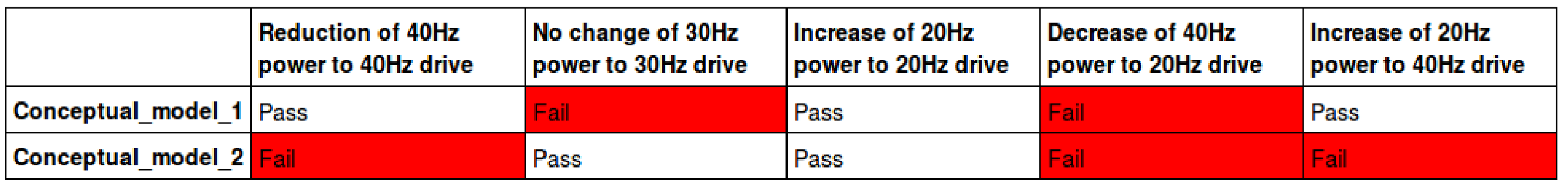}}
\caption{Contrasting the results of comparing two models against experimental observations. a) Model instances are created. b) Appropriate tests are created. c) Tests are grouped together to form a test suite, then the models
are run against the test suite. d) A comparison table shows the performance of each model against each test.}
\label{Fig:ModelComparisons}
\end{figure}

\paragraph{Simple model comparison}
By creating tests, based on the model capabilities, and grouping 
them into test suites, we can easily compare models against experimental data and against each other. Figure \ref{Fig:ModelComparisons} demonstrates how we can use the module to create two different models along with
several tests, then run the models to produce the data relevant for the tests and then judge the model outputs against experimental data and  display the result together. Note that in this context
we use the term model as the \textit{in silico} instantiation of a theoretical/conceptual model. Two different models therefore, do not necessarily have to use different model implementations but might simply differ in parameters.

\begin{figure}
\subfigure[Create model instances and run simulation]
{\includegraphics[width=0.85\textwidth]{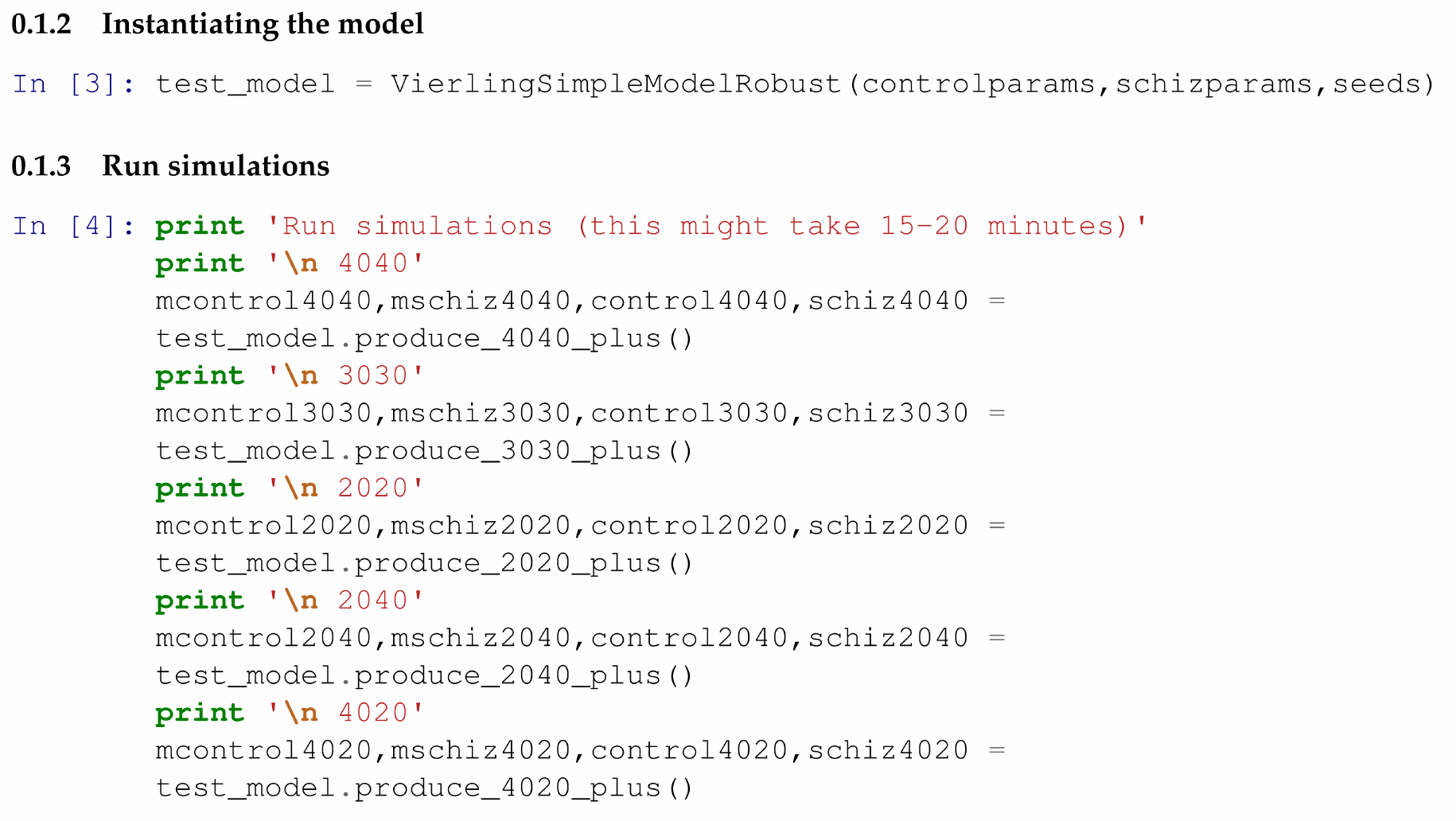}}
\subfigure[Boxplot of model data]
{\includegraphics[width=0.65\textwidth]{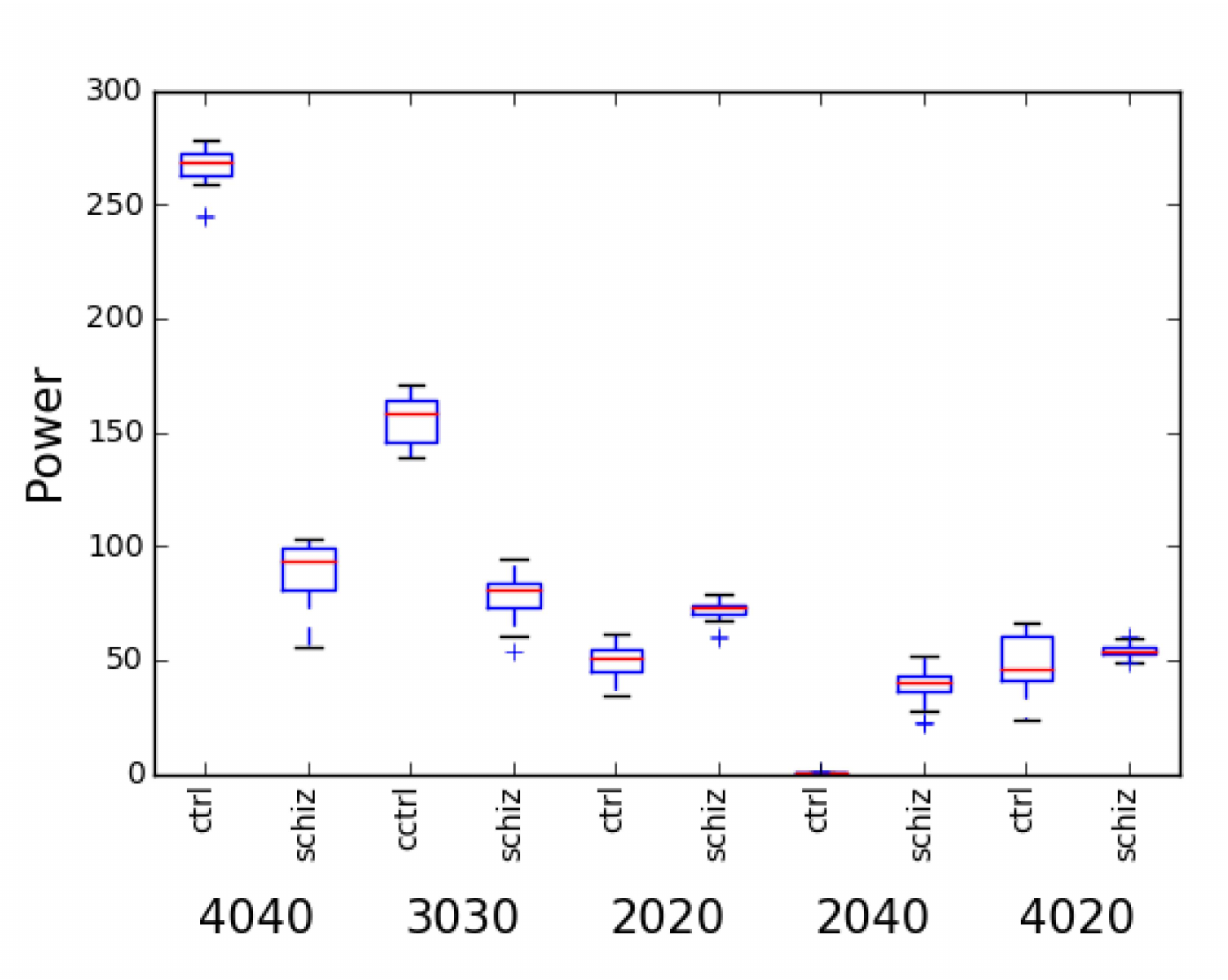}}
\subfigure[Statistical analysis of model data]
{\includegraphics[width=0.85\textwidth]{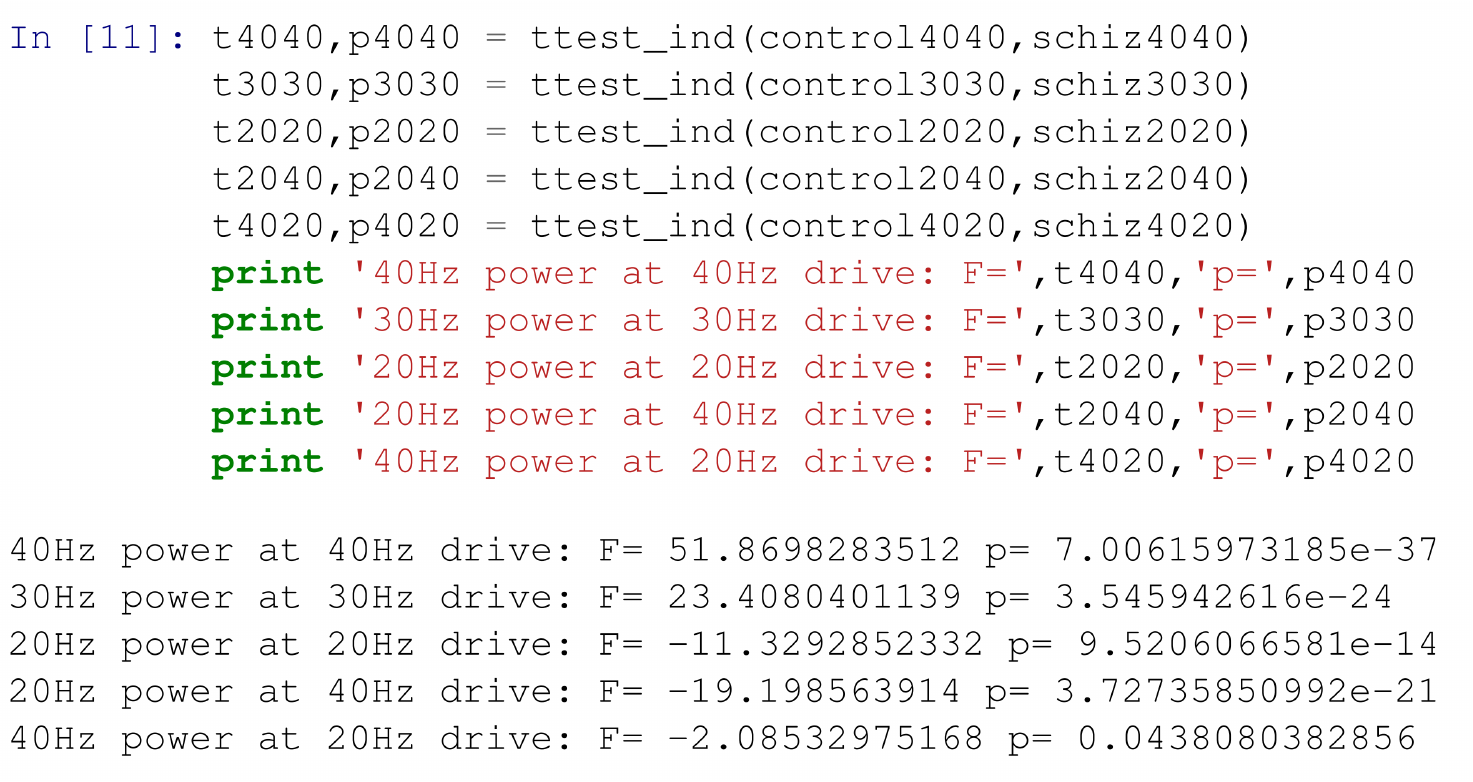}}
\caption{Generating additional data with the '...\_plus' methods of the model classes. a) Model instances are created and the '...\_plus' method is used to run the simulation. b) The additional data visualized
as a boxplot. c) Statistical analysis of the additional model data.}
\label{Fig:ModelPlusMethods}
\end{figure}

\paragraph{Advanced modeling data and visualization}
As already described in the Methods section, the model classes do not only contain the standard methods that implement the necessary capabilities, but also contain so-called '...\_plus' methods 
which generate additional model data. Together with the methods from the visualization class, this additional model data can be used to better understand the model behavior, 
to judge the robustness of findings and to statistically analyze model output (see Figure \ref{Fig:ModelPlusMethods}).

\subsection{Use Case III: Overview over Model Predictions}
Finally, we show how predictions can be generated from existing models (see Figure \ref{Fig:ModelPlusMethods}). In order to generate the predictions, a set of prediction tests along with prediction capabilities, that is, 
capabilities the models must have 
in order for the model to generate the relevant data needs to be created. For demonstration purposes, we have chosen to implement a single, simple prediction test. Since in \textit{ASSRUnit} so far, we have only looked 
at experimental observations and computational models that cover gamma and beta range entrainment, the first test simply generates a prediction how, in a given model, power in the alpha band (here at 10\,Hz)
differs between the control network and the schizophrenia-like network at 10\,Hz drive.
Note that this prediction 
test has been studied in the experimental literature, which means that it could have already been included in the experimental database and therefore does not represent a true prediction.
However, we have chosen to include it for the purpose of demonstration. 
\begin{figure}
\label{Fig:Tests4040}
\subfigure[Create model instances and tests, and run model]
{\includegraphics[width=0.9\textwidth]{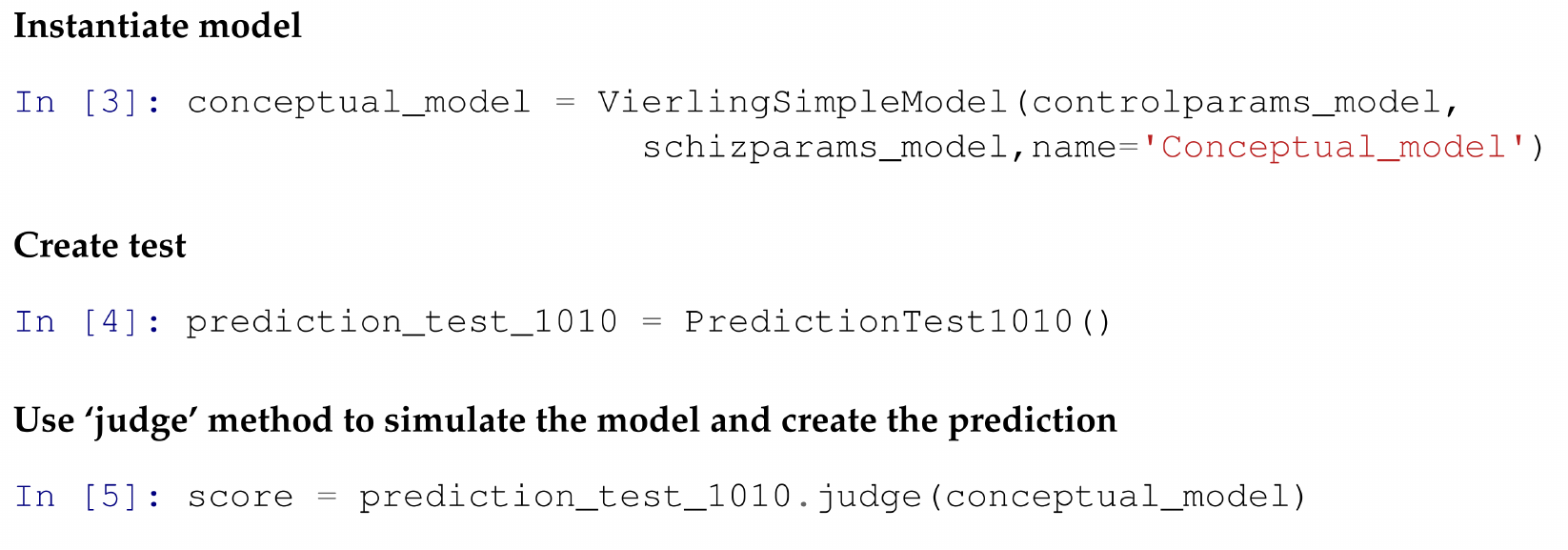}}
\subfigure[Display results]
{\includegraphics[width=0.9\textwidth]{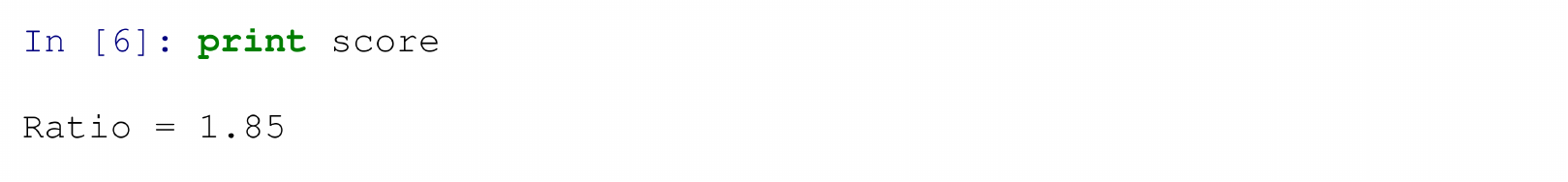}}
\caption{An overview of predictions from a model.}
\end{figure}

\FloatBarrier
\section{Discussion}

\paragraph{The potential role of the framework within computational psychiatry}
The use of computational approaches has seen a significant increase over the last decades in almost all areas of medicine and life sciences. Especially in psychiatry it has 
become clear that the complex and often polygenic nature of psychiatric disorders might only be understood with the help of computational models \cite{Adams2016,Wang2014,Friston2014,Corlett2014,Stephan2014,
Montague2012,Siekmeier2015}.
Naturally, the number of computational models in the field of psychiatry has also increased significantly over the last years and it has been argued that 
\textit{in silico} instantiations of biomarkers and endophenotypes are a crucial step towards an understanding of underlying disease mechanisms \cite{Siekmeier2015}. While this large increase in 
modeling studies shows the importance of computational methods in the field, it also raises several issues that, impede the community to exploit these 
approaches to their full potential. In order for a computational model to be a substantial contribution to knowledge it has to adequately instantiate experimental observations,
correctly implement the mathematical equations of the model and generate experimentally testable predictions. The approach presented here, addresses two of these three requirements, namely, 
the instantiation of experimental observations and the generation of testable predictions. While correctness of the code is an equally important requirement, it was out of scope of the
current work, since it very strongly depends on the type of computational model and on the programming language used to implement the model. Nevertheless, the
approach  presented here offers significant benefits for, not only the computational psychiatry community, but for the psychiatry community as a whole, while imposing little additional efforts
for the users and contributors. It gives modelers a tool to query experimental observations on neurophysiological and neurocognitive biomarkers, and therefore, helps them to
include the current relevant experimental data into their modeling efforts. It further enables them to validate their modeling output against experimental observations during model construction
and to demonstrate the performance of their model, both, with respect to the experimental literature and with respect to other competing models. In addition to the benefits it offers
the modelers, it also enables experimentalists to quickly gain insight into the current state of modeling and to extract experimentally testable predictions from the models. Last but not least,
it offers a tool to reviewers which allows them to judge a newly proposed model by making explicit its performance against experimental data and competing models.

The concept of automated code testing and validation has been successfully applied in computer science for many years now, however, it is only slowly finding its way into the computational 
branches of scientific fields. SciUnit attempts to satisfy this demand by providing a simple, flexible yet powerful framework to address the above-mentioned issues. The computational neuroscience 
community has started to adopt this framework for the automatic validation of single neuron models (NeuronUnit, \cite{Gerkin2014}). 
To the best of our knowledge, we are not aware of any similar efforts in the field of psychiatry.

Since schizophrenia is a polygenic, multi-factorial and very heterogeneous disorder, it has been argued that the usefulness of biomarkers and endophenotypes lies in their potential to
dissect the disorder into subtypes, which might even be linked more closely to findings on the genetic level \cite{Meyer2006,Perlis2011,Markou2009}. The proposed \textit{ASSRUnit} module together with computational models of 
biomarkers/endophenotypes and specifically designed test suites could strongly facilitate this process by providing mechanistic links between neurophysiological or neurocognitive biomarkers
and changes at the synaptic,cellular and/or network level.

\paragraph{Future directions for ASSRUnit}
The presented \textit{ASSRUnit} module can be easily extended and modified by others to fit their needs (for example to include  more specialized visualization tools). Our efforts
for establishing \textit{ASSRUnit} as a widely used tool will focus on three main areas: 1) We aim to cover the majority of existing experimental studies with our experimental database in the future.
Furthermore, we hope to convince experimentalists to provide more detailed experimental data or to
ideally create database entries themselves. 2) We also aim to cover the majority of current computational models that describe the cortical circuitry responsible for the ASSR. 
Again, we hope to encourage modelers to actively contribute to \textit{ASSRUnit}.
3) We aim to extend our set of prediction tests, and thus, our prediction database. 

The most straightforward extension, in our view, is to include information on phase-locking in addition to 
pure power in certain frequency bands. 
Several studies report, additionally to a reduction in gamma power, a  reduction in the phase-locking factor for patients suffering from schizophrenia 
(for example \cite{Kwon1999,Brenner2003,Light2006,Vierling2008,Krishnan2009}). These observations can very easily be incorporated into the existing module, simply by including the experimental observations into the database,
adding the necessary capabilities to the model classes and by adding the appropriate tests that link the experimental observations to the model capabilities.

Furthermore, the changes in oscillatory activity upon auditory stimulation are not limited to the gamma and the beta range for schizophrenic patients, but also extend to lower frequency bands such as alpha, theta and delta.
For example, Brockhaus-Dumke and colleagues find reduced phase-locking in the alpha and theta band for schizophrenic patients in an auditory paired-click paradigm \cite{Brockhaus2008}, and Ford et al. find a
reduction of phase-locking in the delta and theta range for schizophrenic patients in an auditory oddball task \cite{Ford2008}. Abnormalities in these frequency bands have also been found in many other paradigms outside of
the auditory system (see \cite{Basar2013}). To the best of our knowledge, ASSRs to entrainment stimuli in the theta and delta range have not been looked at in schizophrenia. Therefore, \textit{ASSRUnit} could be 
used to generate predictions in these frequency ranges as demonstrated in use case III.

However, an inclusion of the above-mentioned observations together with computational models explaining these deficits is not straightforward, because either the paradigms are different from the ones
used to elicit ASSRs and/or the mechanisms underlying the effect are different, and therefore the computational models, are substantially different to models of ASSRs. Therefore, these deficits are better explored in separate modules solely focusing
on each paradigm/deficit. However, it would be very interesting to 'co-explore' computational models that have the capabilities to explain both, ASSR gamma/beta band and delta/theta/alpha phase-locking, deficits. Such an analysis 
could highlight interactions between different mechanisms underlying different symptoms/biomarkers.

Another very interesting and promising extension of the current module would be to include data and models from different psychiatric disorders, since schizophrenia is not the only disorder
where patients show entrainment deficits. Wilson et al. \cite{Wilson2007}, explored gamma power adolescents with psychosis and found reductions compared to normally developing controls. Their patient group
consisted of patients suffering from schizophrenia and also from schizoaffective disorder and bipolar disorder. Interestingly, these disorders show overlapping symptoms, neurobiological substrates and predisposing gene loci. 
Other studies have also found reduced power and phase-locking in the gamma range in patients 
with bipolar disorder \cite{ODonnell2004,Spencer2008,Rass2010}.
The presented module is perfectly suited to highlight commonalities and differences across disorders and to link those to mechanistic 
explanations via different theoretical/computational models.

\paragraph{Other modules beyond ASSRUnit}
The approach presented here, combining an experimental database with a collection of models, tests, prediction tests and a resulting predictions database, can be readily applied 
to a number of other neurophysiological biomarkers of schizophrenia as well as other psychiatric disorders. 
In patients suffering from schizophrenia a dysfunction of the auditory system has long been suspected. In fact, a large number of biomarkers and endophenotypes for schizophrenia, other
than ASSR deficits, involve auditory processing. Several alterations of event-related potentials (ERPs) such as mismatch negativity (MMN), N100, and P50 have been described in the literature 
(see also \cite{Siekmeier2015,Shi2007}).

MMN is a negative component of the auditory evoked potential, which is evoked by an alteration in a repetitive sequence of auditory stimuli. MMN
seems to be specific to schizophrenia because patients suffering from other psychiatric disorders (for example bipolar disorder and major depression) show normal MMN \cite{Umbricht2003}. 
Auditory MMN is likely to be generated in primary and secondary auditory cortices and is therefore very similar  
to stimulus-specific adaptation (SSA) properties of single neurons in auditory cortex \cite{Nelken2007} (although not identical; see efor example \cite{Farley2010,vonderBehrens2009}).
Several models explaining mechanisms underlying MMN/SSA have been proposed (for example \cite{Mill2011,Nelken2014}).  

The P50 potential, a small positive deflection of the EEG signal at around 50\,ms after the onset of an auditory stimulus, is often reduced to the second of a pair of stimuli.
However, this so-called P50 reduction is markedly reduced in schizophrenic patients (for example \cite{Braff2007}). Another important measure of sensory gating is the pre-pulse inhibition (PPI) of
the auditory startle reflex (i.e. the phenomenon in which a weaker prestimulus inhibits the reaction to a subsequent strong startling stimulus). As P50 reduction, PPI is also reduced in schizophrenic patients,
although these phenotypes do not seem to correlate \cite{Braff2007}. Again, computational models exploring the mechanisms underlying PPI have been developed (for example \cite{Schmajuk2005,Ramirez2012,Leumann2001}, and Moxon et al.
have investigated the dopaminergic modulation of the P50 auditory-evoked potential and its relationship to sensory gating in schizophrenia \cite{Moxon2003}.
 
The N100 (also called N1) is a negative component of the EEG signal occurring approximately 100\,ms after stimulus onset. This negative deflection is again reduced in schizophrenic patients 
(reviewed in \cite{Rosburg2008,Rissling2010,Javitt2008}) and these deficits have also been modeled \cite{Ventouras2000}).

Although we here concentrated on biomarkers and deficits in the auditory cortex, our approach is well adaptable to brain circuits outside of the auditory system. 
Working memory deficits are probably one of the most robust and best described cognitive deficits in schizophrenic patients (reviewed in \cite{Piskulic2007,Lee2005}). Patients
show a decrease in working memory capacity, i.e. the capacity to maintain, manipulate and use information online for a relatively short period of time, across a broad range of paradigms.
Again, several theoretical and computational models have been proposed, aiming to provide mechanistic descriptions of the underlying mechanisms (for example \cite{Compte2000,Durstewitz2000,Wang2004,Singh2006,Wang2001,Cano2012}). 

All these deficits and alterations along with the mentioned computational models could be integrated into a module similar to the proposed ASSRUnit module.
Such a unified framework would be of great benefit for the study of schizophrenia pathology due to the diversity of symptoms, biomarkers, and experimental observations linked to the mental disease.

\section{Conclusion}
We have proposed a framework for automated validation and comparison of computational models of neurophysiological and 
neurocognitive biomarkers of psychiatric disorders. The approach builds on
SciUnit, a Python framework for scientific model comparison. As case in point, we used this framework to develop \textit{ASSRUnit}, a module comprising 
an experimental observations data base, computational models, capabilities, tests/test suites and visualization functions for
ASSR response deficits in schizophrenia. 

Our approach will facilitate the development, validation and comparison of computational models of neurophysiological 
and neurocognitive biomarkers of psychiatric disorders by making the scope of models explicit and by 
making it easy for the user to assess a model's validity and to compare a model against competing models. Furthermore, it is easy to use, 
straightforward to extend to more experimental observations, computational models and analyses and,
ready to apply to other biomarkers. Therefore, the adoption of the proposed framework could be of great use for modelers, reviewers and 
experimentalists in the field of computational psychiatry.

\bibliographystyle{plain} 
\bibliography{mybib}

\end{document}